\documentclass[aps,12pt]{revtex4}
\usepackage{amssymb}
\usepackage{graphicx}
\usepackage{dcolumn}
\usepackage{bm}

\begin{document}

\title{Measurement of spatial distribution of cold atoms in an integrating sphere}
\author{Xu-Cheng Wang}
\author{Hua-Dong Cheng}
\author{Ling Xiao}
\author{Yan-Ling Meng}
\author{Liang Liu$^*$}
\author{Yu-Zhu Wang}

\affiliation{Key Laboratory of Quantum Optics, Shanghai Institute
of Optics and Fine Mechanics, and Center for Cold Atom Physics,
Chinese Academy of Sciences,
Shanghai 201800, China\\
$^*$Corresponding author: liang.liu@siom.ac.cn}

\date{\today}

\begin{abstract}
In this paper, we present an experiment to measure the spatial
distribution of cold atoms in a ceramic integrating sphere. An
quadrupole field is applied after the atoms are cooled by diffuse
light produced in the ceramic integrating sphere, thus the shift of
atomic magnetic sub-levels are position-dependent. We move the
anti-Helmholtz coil horizontally while keeping the probe laser beam
resonant with the cold atoms at the zero magnetic field. The
absorption of the probe beam gives the number of cold atoms at
different position.  The results show that at the center of the
integrating sphere, less atoms exist due to the leakage of diffuse
light into the hole connecting to the vacuum pump. The method we
developed in this paper is useful to detect cold atoms in a region
where imaging is not possible.


\end{abstract}

\maketitle


Cooling atoms by diffuse light has received a lot of attention
recently due to its important application in compact cold atom
clock, the HORACE for example
\cite{guillot99IEEE_horace,tremine04IEEE_horace,tremine05IEEE_limitations,esnault07IEEE_stability,esnualt08FSM_reaching}.
The HORACE has reached $2.2\times 10^{-13}\tau^{-1/2}$ short term
stability \cite{esnualt08FSM_reaching}, and is expected to be
further improved in microgravity environment
\cite{tremine05IEEE_limitations}. Cooling atoms from a background
vapor in a glass cell by diffuse light was first realized in
cesium in 2001 \cite{guillot01OL_three}, and then in rubidium in
2009 \cite{cheng09PRA_laser}. Typically, diffuse light is produced
by multiple reflection of laser light at the inner surface of a
hollow sphere with high-reflective medium which encloses the glass
cell. Such a device is similar to an integrating sphere, which is
normally used for photometric or radiometric measurements.

It is quite obvious that in a completely-enclosed sphere, if the
reflectivity of the inner surface is high enough, the light field in
the sphere should be homogeneous. The distribution of cooled atoms
in a homogeneous diffuse laser field should be homogeneously
distributed in the region where the diffuse field exists. But in
fact, the light field in the sphere is not homogenized due to
several reasons, including the injection of laser light, hole in the
sphere for the vacuum pump, inhomogeneous reflectivity of the inner
surface of the sphere, and et.al. Such an inhomogeneity leads to
inhomogeneous distribution of cooled atoms in the sphere. Esnault
{\it et al.} suggested that the shape of cold atom cloud in the
sphere has two lobes due to the leakage of diffuse light at two
holes used to connect the vacuum pump and atom reservoir
\cite{esnault07IEEE_stability}, but no convincing results have been
shown, and no direct measurement of the shape has been done so far
due to the difficulties to measure the shape in a closed hollow
sphere, in which imaging of fluorescence of cold atom cloud is not
possible. Measuring and reshaping the cold atom cloud are very
important works for the compact cold atom clock, and thus it is
necessary to measure the distribution of cold atoms in the sphere.

\begin{figure}
\centering
\includegraphics[width=0.3\textwidth]{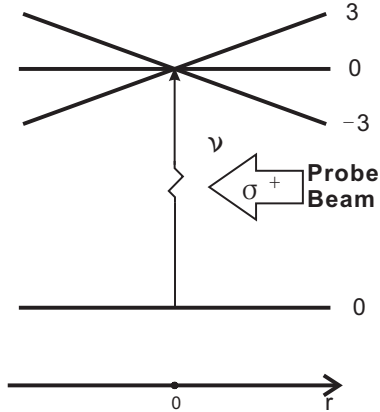}
\renewcommand{\figurename}{Fig.}
\caption{\label{principle}The schematic diagram for the principle of
the method.}
\end{figure}

In this work, we developed a method to measure the spatial
distribution of cold atoms in a closed sphere. The principle is
similar to that of magneto-optic trap (MOT), as shown in Fig.
\ref{principle}.
In the MOT, an anti-Helmholtz coil is used to create a quadrupole
magnetic field with zero at the center. With a pair of
counter-propagating $\sigma^+-\sigma^-$ laser lights, the force on
an atom is position-dependent because of the position-dependent
energy shift of the atomic magnetic sub-levels, and thus atoms in
the MOT can be cooled and trapped \cite{metcalf99Book_laser}. In our
case, the atoms are pre-cooled by diffuse light
\cite{guillot01OL_three,cheng09PRA_laser}. The cooled atoms at
different positions encounter different magnetic fields, and thus
different energy shifts. By moving the anti-Helmholtz coil while
keeping the probe laser resonant with the atomic transition at zero
magnetic field, we can map the spatial distribution of cold atoms in
the cell by recording the absorption signal of the probe laser. Here
we give one dimensional case.

For $^{87}$Rb, the ground states have sub-levels $F=1,2$, and
excited states $F'=1,2,3$. A $\sigma^+$-circularized probe laser
beam, resonant with the transition of $F=2\rightarrow F'=3$, is used
to detect the cold atoms around the region of zero magnetic field.
Atoms aside this region are not detected due to the Zeeman
splitting.

\begin{figure}
\centering
\includegraphics[width=0.4\textwidth]{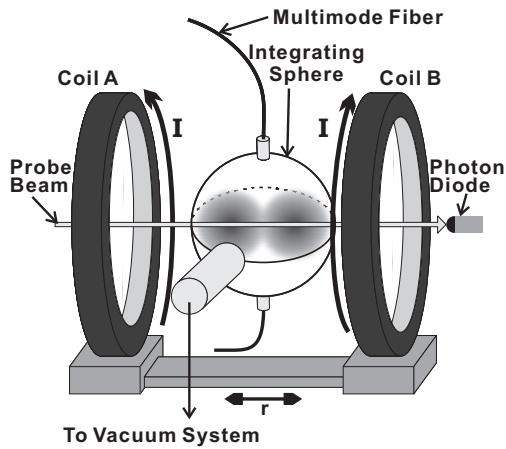}
\renewcommand{\figurename}{Fig.}
\caption{\label{setup}A schematic of the experimental setup. The
cooling and repumping lights are coupled into the integrating sphere
vertically with two multimode fibers.}
\end{figure}

The experimental setup was discussed in
Refs.~\cite{cheng09PRA_laser,zhang09PRA_observation}, here we
briefly introduce the setup as shown in Fig. \ref{setup}. A glass
cell, connected to a vacuum pump through a glass tube with inner
diameter of 8 mm, is filled with rubidium vapor at a background
pressure around $10^{-7}$ Pa. The glass cell has an inner diameter
of 43 mm, and is surrounded by an integrating sphere made of ceramic
whose diffuse reflection coefficient at the inner surface is 98\% at
780 nm. Two laser beams, containing both cooling and repumping
lights, are injected into the integrating sphere through two
multimode fibers vertically. With multiple reflection, diffuse light
is formed in the sphere. When the cooling laser is tuned at
appropriate frequency, the atoms can be cooled by diffuse light
\cite{guillot01OL_three,cheng09PRA_laser}.

\begin{figure}
\centering
\includegraphics[width=0.5\textwidth]{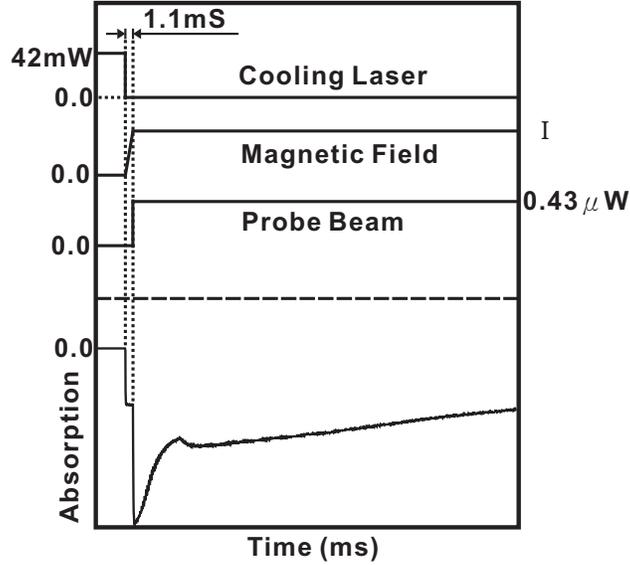}
\renewcommand{\figurename}{Fig.}
\caption{\label{sequence}The experimental sequence with a typical
results in the bottom.}
\end{figure}

Figure \ref{sequence} gives the typical experimental sequence and
result. First, the cooling laser is turned on for 6 s, which is long
enough for the loading of cold atoms \cite{xiao09COL_analysis}. Then
the magnetic field is turned on right after the cooling light is
switched off. After 1.1 ms, when the magnetic field becomes stable,
 the horizontal probe beam is switched on. Since the
lifetime of cold atoms in the glass cell can be longer than 40 ms
\cite{zhang09CPL_lifetime}, the change about the spatial
distribution with 1.1 ms delay can be neglected. In the experiment,
the power of the injected cooling laser into the integrating sphere
is 42 mW, and the cooling laser is red detuned 18 MHz (about
$-3\Gamma$) to the transition of 5$^{2}S_{1/2}$, $F=2 \rightarrow
5^{2}P_{3/2}$, $F=3$, with which the diffuse light has the top
efficiency in capturing the $^{87}Rb$ atoms. The power of the
repumping light is about 5 mW, which is locked to the resonant
transition of 5$^{2}S_{1/2}, F=1 \rightarrow 5^{2}P_{3/2}, F=2$.
\begin{figure}
\centering
\includegraphics[width=0.5\textwidth]{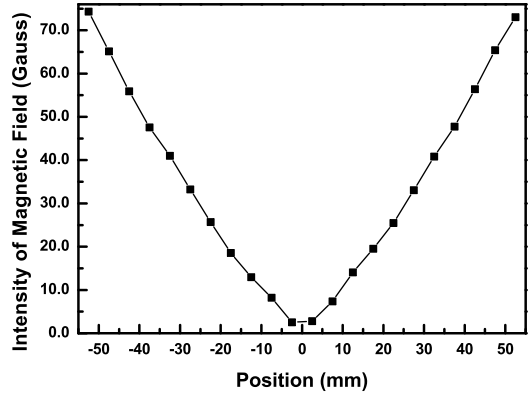}
\renewcommand{\figurename}{Fig.}
\caption{\label{mag13gs}The gradient is 1.30 Gauss/mm.}
\end{figure}

The bottom of Fig.~\ref{sequence} gives a typical absorption signal,
where the gradient of the magnetic field is around 1.30 Gauss/mm, as
shown in Fig.~\ref{mag13gs}, and the zero-point of the magnetic
field is about 12 mm from the center of the integrating sphere. The
spatial distribution can be determined from the absorption signals
by moving the anti-Helmholtz coil along the probe beam with the step
of 5 mm.

\begin{figure}
\centering
\includegraphics[width=0.5\textwidth]{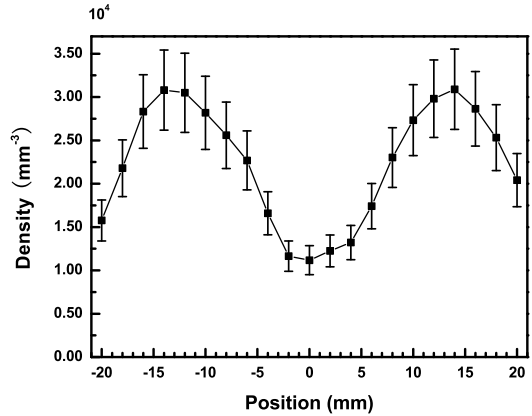}
\renewcommand{\figurename}{Fig.}
\caption{\label{atom13gs}The density of the cold atoms vs position
when the magnetic field gradient is 1.30 Gauss/mm.}
\end{figure}

From the absorption of the probe beam we can work out the density of
the cold atoms at different positions, as shown in
Fig.~\ref{atom13gs}. Clearly, the spatial distribution shows that at
the center of the sphere, less atoms exist, and the atoms are mostly
distributed at the position around $r=\pm$14 mm from the center of
the integrating sphere (where $r=0$). The density of the cold atoms
at $r=-14$ mm is about 3.08 $\times$ 10$^{4}$ $mm^{-3}$, while at
$r=0$ mm, is about 1.18 $\times$ 10$^{4}$ $mm^{-3}$. This is mainly
because the hole of the glass tube, which is connected to the vacuum
pump, as discussed in Ref. \cite{esnault07IEEE_stability}

The position resolution of our measurement is mainly determined by
the spontaneous emission rate of the excited state and the linewidth
of the laser. The magnetic field along the probe beam created by the
anti-Helmholtz coil can be simply described as
B($\Delta$r)=c$\Delta$r, here, c is a constant, and $\Delta$r is the
position with origin at the zero-point of the quadrupole field. For
weak magnetic field, the energy levels are split linearly according
to
\begin{equation}
\Delta E_{Zeeman} = g_{F}\mu_{B}m_{F}B(\Delta r) =
g_{F}\mu_{B}m_{F}c\Delta r \label{eq:remainatoms}
\end{equation}

As the probe beam is turned on, most of the atoms in the $F=2$,
$m_{F}=-2,-1,0,1$ states can be quickly pumped into $F=2$,
$m_{F}=2$, and connecting $F^{'}=3$, $m_{F^{'}}=3$, such a
transition system becomes closed circles. And besides that, the
Clebsh-Gordan coefficient $F=2, m_{F}=2 \rightarrow F^{'}=3$,
$m_{F^{'}}=3$ transition is $\sqrt{\frac{1}{2}}$, which is much
larger than the other transitions of $F=2$ $\rightarrow$ $F^{'}=3$.
So we can only consider $F=2$, $m_{F}=2 \rightarrow F^{'}=3$,
$m_{F^{'}}=3$.

The Zeeman shift of $F=2$, $m_{F}=2$ is 2 $\times $ 0.70 MHz/Gauss,
and $F^{'}=3$, $m_{F^{'}}=3$ is 3 $\times$ 0.93 MHz/Gauss. Thus from
Eq.(1) the resonance of probe beam with the $F=2$, $m_{F}=2$
$\rightarrow$ $F^{'}=3$, $m_{F^{'}}=3$ transition is
\begin{equation}
\Delta\nu=(3 \times 0.93 - 2 \times 0.70)B(\Delta r)=1.39c \Delta r
\end{equation}
so we get
\begin{equation}
\Delta r = \frac{\Delta\nu}{1.39c} \label{eq:remainatoms}
\end{equation}
Here, $\Delta$r can be defined as the position resolution.
Obviously, the larger gradient of the magnetic field, the higher
position resolution.

The natural linewidth of $^{87}$Rb is about 6 MHz, and the linewidth
of the laser is about 1 MHz, suppose $\Delta\nu=7$ MHz, so
\begin{equation}
\Delta r = \frac{\Delta\nu}{1.39c} = \frac{5.04}{c}
\label{eq:remainatoms}
\end{equation}

For $c=1.30$ Gauss/mm, $\Delta r=3.88$ mm. The resolution is high
enough to determine the spatial distribution of cold atoms in the
sphere, which has a 43 mm diameter.

In conclusion, we have developed a method to measure the spatial
distribution of cold atoms in a nearly enclosed sphere where imaging
is not possible. We pointed out that the spatial resolution is as
high as 3.88 mm in our measurement. With the method, we measured the
spatial distribution of cold atoms in an integrating sphere, and
found that the distribution has two lobes. This result is very
important for developing a compact cold atom clock.

Supports from the National Nature Science Foundation of China
under Grant No. 10604057 and 10874193, National High-Tech
Programme under Grant No. 2006AA12Z311 and the National Basic
Research Programme of China under Grant No. 2005CB724506 are
gratefully acknowledged.

\bibliography{Integrating_ref}

\end{document}